# On Generation of Firewall Log Status Reporter (SRr) Using Perl

Sugam Sharma[1], Hari Cohly[2], and Tzusheng Pei[2]

[1]Department of Computer Science, Iowa State University, USA
sugam.k.sharma@gmail.com
[2]Center for Bioinformatics, Jackson State University, USA
{hari.cohly, tzusheng.pei}@jsums.edu

*ABSTRACT*

*Computer System Administration and Network Administration are few such areas where Practical Extraction Reporting Language (Perl) has robust utilization these days apart from Bioinformatics. The key role of a System/Network Administrator is to monitor log files. Log file are updated every day. To scan the summary of large log files and to quickly determine if there is anything wrong with the server or network we develop a Firewall Log Status Reporter (SRr). SRr helps to generate the reports based on the parameters of interest. SRr provides the facility to admin to generate the individual firewall report or all reports in one go. By scrutinizing the results of the reports admin can trace how many times a particular request has been made from which source to which destination and can track the errors easily. Perl scripts can be seen as the UNIX script replacement in future arena and SRr is one development with the same hope that we can believe in. SRr is a generalized and customizable utility completely written in Perl and may be used for text mining and data mining application in Bioinformatics research and development too.*

## 1. INTRODUCTION

Practical Extraction Reporting Language (Perl) [1] has attained enormous popularity in a very short duration in different computational areas such as Bioinformatics due to easy use of its vast open source library available. Perl is not only helpful in Bioinformatics development but also in core computer areas at the corporate level. Computer System/Network Administrator generates lot of reports to get the status of different files available. More involvement is of log files which are generated very frequently and tracking is utmost important to resolve the involved issues. To scan the summary of large log files and to quickly determine if there is anything wrong with the server or network, Firewall Log Status Reporter (SRr) helps to generate the reports day to day basis based on the parameters of interest. SRr provides the facility to administrators to generate the individual firewall report or all reports in one go.

SRr needs a flat text input file to operate on. The contents of the input file should be separated by some kind of separator e.g. white space, comma, semicolon, full stop etc. SRr reads the input file line by line and stores the contents of that line in an array. Subsequently the contents of that array are stored in scalar variables for further use. SRr is a customized tool and depending of one's requirement can be modified. In our development we use a firewall log file in textual format and one of its attributes is message type. This attribute is of our great interest as we deal with log files in our operation and we collect only that data from input file whose message type is log. Though there are many attributes a file consists of, but few of them are attributes of interest, we consider for generating the status report. In our work we consider fours attributes src (source), dst(destination), service, and s_port as attributes of interest out of a big bunch of attributes the firewall log input file contains. Based on these attributes we generate four different reports, Source addresses report, Destination addresses report, Service usage report, Network interface report. These reports can be generated individually or all at a time.

Generation of source addresses report by SRr is based on source attribute of input file. All the source addresses in firewall log file under log message type are collected in an array. It depends on individual's requirement what kind of status report it needed and based on one's need the contents of that array are extracted and utilized. For our requirement we want to display the status report which consists of the name of the source address, the frequency of request made by this source, and the percentage of request made by this source. As log files are pretty long and updated very frequently, there may be a possibility that a generated status report has a long list of sources and difficult to analyze even after pagination. So to analyze the status report efficiently we sort the generated report based on percentage the request made in decreasing order.

SRr generates the destination addresses report using destination attribute of the input file. All elements of destination attribute in log file under log message type are collected in an array. SRr uses this array to generate the status report in the same way as we explain above for source attribute.

Based on our requirement we use a composite attribute consisting of protocol and service primitive attributes of input log file. SRr uses this composite attribute to generate the status report. We extract those two primitive attributes individually from the input file and combine them together and store in a user defined composite attribute. All elements of the composite attributes are stored in a user defined composite array. This is the array where SRr operates on to generate the status report in the same fashion as explained above for source attribute. To generate the network interface status report SRr uses a user defined composite attribute. The composite attribute is produced by the combination of interface name attribute, interface directory attribute, and protocol attribute of input text file we use for our purpose. The composite attribute is further classified based on inbound and outbound request factors of interface _dir primitive attribute of input text file. The functioning of SRr on this attribute to generate the report is similar to that stated above for source attribute. The rest of the paper is organized as follows.

Section 2.0 is the design section. This section details about the major contents of the tool and its working. Each subsection describes about individual content of the tool in detail. Section 3.0 is the result section. Section 4.0 concludes the paper. The last section 5.0 is the reference section.

## 2.0 DESIGN

Firewall Log Status Reporter has three major components. The very first and essential component is the input file. The only restriction that has been imposed on input file is that it should be a flat file in textual format. The contents of the file should be separated by some kind of separator such as white space, comma, semicolon, full stop etc. SRr opens the input file using a file hander

(*open ALLREPORTS, "<$prc_file" || die "Could not open file for reading.\n"* )

and reads the opened file one line at time. During the reading of a line, SRr splits the line contents and stores the values in an array. In Perl split is the function which helps in splitting (*my@cur_line = split /:/, $*). According to the Perl documentation if there is no explicit declaration of an array, *@_* is the default array which stores the values and *$_* is the default variable which stores a scalar value.

The next step is to assign the array values to different variables for further use. We use shift function [2] to serve this purpose. The use of shift function very first time on an array moves the very first element of that array to the left hand side variable. The use of next consecutive shift function moves the second element of the array to the left hand side scalar variable. Like wise we proceed and store all the array elements to the different scalar variables for further operation.

SRr is a customized tool and in our development we collect only those lines in which the message type attribute is log (*$type_of_message eq "log").* In most of the cases the type of message is log but depending on individual's interest the type of collection can vary.

It is recommended that each status report should contain the date the report has been generated on. For that we use an external date module which generates the current date and associate that with the header of the report generated.

*my @day_name = ("Sun", "Mon", "Tue", "Wed", "Thu", "Fri", "Sat");*
*my @month_name = ("January", "February", "March", "April", "May", "June", "July", "August", "Sept", "October", "November", "December");*
*my ($sec, $min, $hour, $mday, $mon, $year, $wday); ($sec, $min, $hour, $mday, $mon, $year, $wday, undef, undef)=localtime(time()); $year+=1900;$mon++;*
*$report_date=sprintf("%s %s %s %02d:%02d:%02d",$day_name[$wday],$month_name[$mon], $year, $hour, $min, $sec);*
*print LOGREPORTS "Report generated on:$report_date \n".*

In this above code snippet LOGREPORTS is the output handler which writes the data in user defined output file.

There are numerous attributes in the text file that has been read. It depends on the individual's interest and requirement which attributes are used as attribute of interest to generate the status report. In our work we consider four attributes in input text file of our interest and generate the status reports based on them. These attributes are src (source), dst(destination), service, and s_port. Based on these attributes we generate four different reports. We provide a flexibility to generate a single status report as a whole consisting of all individual reports. As mentioned earlier, we use modular programming in SRr development. Corresponds to each individual report a subroutine has been defined. Each subroutine governs the functioning of status report generation for SRr against one parameter of interest.

## 2.1 SOURCE_ADDRESSES_REPORT

This section describes the functioning of the SRr to generate the status report based on source attribute of input file. In the following block of code *@src_array* is a filled array consisting of the source elements of log type messages of input file. We traverse the array in foreach loop. Though it depends on individual's requirement what kind of status report it needs. For our requirement we want to display the status report which consists of the name of the source address, the frequency of request made by this source, and the percentage of request made by this source. There may be several sources which attempt to enter the network and succeed. As the larger percentage of occurrence contributes more significantly in analysis of status report so we sort the report based on percentage in decreasing order in report display. In the following segment of code

*for (@src_array) { $r++ if /$item/i;*
*}*

is responsible for the calculation of frequency of occurrence of request from a specific source in a network. The fraction in following block sorts the array in decreasing order.

*sub Source_Addresses_report{ foreach $item (@src_array)*
*{*
*for (@src_array) { $r++ if /$item/i;*
*}*
*$percent = sprintf("%.2f", ($r/$count)*100); $item_1 = substr $item, 0, 16;*

*$string = $percent." ".$item_1." \t".$r." \t".$percent."%" ; if ( !grep( / $string/,@string_array ) ) { $max_report_entries++;*
*push(@string_array,$string);*
*}*
*$r=0;*
*}*
*foreach $item_sort ( reverse sort @string_array)*
*{*
*$item_sort_sub = substr $item_sort, 5;*

*print LOGREPORTS " $item_sort_sub \n"; $max_report_entries_1++;*

*}*
*print LOGREPORTS "\n<-----------------Top $max_report_entries_1 of $max_report_entries Entries----------*
*-->\n\n"; $max_report_entries=0; $max_report_entries_1=0; }*

## 2.2 DESTINATION_ADDRESSES_REPORT

This section describes the functioning of the SRr to generate the status report based on destination attribute of input file. In the following block of code *@dest_array* is a filled array consisting of the source elements of log type messages of input file. The description of the code in this section is resembled to that of Source_Addresses_report section except that we use *@dest_array*. If this block of code is chosen to generate the status report, the generated report will display the name of destination, the frequency of request sent to this destination, the percentage of occurrence of request to this destination.

*sub Destination_Addresses_report{ foreach $item (@dest_array)*
*......................*
*……………..*
*……………..*

To avoid the redundancy of code, we replace the duplicate code by dotted line in the above code block for writing purpose.

## 2.3 SERVICE_USAGE_REPORT

This section describes the functioning of SRr to generate the status report based on protocol (proto) attribute and service attributes of input file. In the following block of code *@protocol_service* is a filled array consisting of *$proto_service*. *$proto_service* is the composite attribute of protocol and service attributes (*"$protocol"."__"."$service"*) of log type messages of input file. The status report generated by this section displays the name of the protocol combined with service used by it, the frequency of the usages of the service, and the percentage of usage of the service. This report helps the System/Network administrator to analyze the protocol used and the service used by it.

*sub Service_Usage_report{ foreach $item (@protocol_service)*
*……………*
*……………*
*……………*

## 2.4 NETWORK_INTERFACE_REPORT

This section describes the functioning of SRr to generate status report based on interface name attribute, interface directory attribute and protocol attribute of input text file. *@intf_array* is an array filled by *$intf_name_dir* composite attribute of log type message of input file. *$intf_name_dir* attribute consists of *$originate_devices*, *$interface_name*, and *$interface_dir* concrete attributes of input text file. The status report generated in this section displays the composite attribute name, frequency of involvement of the interface, percentage. Inbound and outbound are two main factors of interface _dir primitive attribute. Based on the report generated System/Network administrator can analyze the origin of request, interface associated and whether the request is inbound or outbound.

*sub Network_Interface_report{ foreach $item (@intf_array)*
*......................*
*……………..*
*……………..*

## 2.5 ALL_REPORTS

This section is important when the administrator is keen to display all status reports in one go. When this segment of code is encountered in processing, the status reports are started to be generated one by one. We try to implement modular programming where ever possible. In this section under the subroutine All_reports, we call different subroutines. Each subroutine represents a module.
This section is important to consider if administrator wants a quick look on all reports to analyze.

*sub All_reports{ Source_Addresses_report(); print".\n";*
*Destination_Addresses_report();print".\n"; Service_Usage_report();print".\n";*
*Network_Interface_report();print"\n";*
*}*

## 2.6 HELP_SCREEN

This section is the help section. SRr requires a command line argument to select a type of status report to be generated. This is user's responsibility to supply that command line argument based on her/his requirement. In the following code snippet we provide six options available for user. We provide a helping hand to the user, if a user misses to supply the command line argument, the help screen will be displayed always.

*sub Help_Screen{*
*print" You need to pass command line argument: Following options are available \n\n".*
 *"s - Create the Source Addresses report\n".*
 *"d - Create the Destination Address report\n".*
 *"u - Create the Service Usage report\n".*
 *"i - Create the Network Interface report\n".*
 *"a - Create all reports\n".*
 *"h - Display the help screen";*
*}*

We use switch statement to switch to different section based on the command line input supplied. The below code segment is self explanatory.

```
switch ($ARGV[0]) {
    case "s"  { Source_Addresses_report()}
    case "d"    { Destination_Addresses_report()}
    case "u"    { Service_Usage_report()}
    case "i"  { Network_Interface_report()}
    case "a"    { All_reports()}
    case "h"    { Help_Screen()}
    else    { Help_Screen()}
  }
```

## 3.0  RESULTS

## 3.1    Input file snippet

In this section we show the curtailed sample input text file we use for our development. The first line is the attribute name and rest of the lines represents the values of these attributes.

*num;date;time;orig;type;action;alert;i/f_name;i/f_dir;proto;src;d  t;service;s_port;len;rule;icmp-type;icmp-code;h_len;ip_vers;sys_msgs*

*1;20Oct2006;17:30:36;192.1.28.3;control;ctl;;daemon;inbound;;;;;;;;;;;;started sending log to localhost*

*2;3Nov2006;13:13:53;192.1.28.3;control;ctl;;daemon;inbound;;;;;;;;;;;;started sending log to localhost*

*3;3Nov2006;13:52:59;fwfoomain01.foo.com;control;ctl;;daemo n;inbound;;;;;;;;;;;;started sending log to localhost*

*4;3Nov2006;14:11:40;fwfoomain01.foo.com;control;ctl;;daemo n;inbound;;;;;;;;;;;;started sending log to localhost*

*5;3Nov2006;14:17:59;fwfoomain01-2;control;ctl;;daemon;inbound;;;;;;;;;;;;;started sending log to localhost*

*6;4Nov2006; 8:42:14;fwfoomain01.foo.com;control;ctl;;daemon;inbound;;;;;
        ;;;;;started sending log to localhost*

*17;17Nov2006;14:02:59;fwfoomain01.foo.com;control;ctl;;dae mon;inbound;;;;;;;;;;;;started sending log to localhost*

*18;17Nov2006;14:03:01;fwfoomain01.foo.com;control;ctl;;lo0;i nbound;;;;;;;;;;;;installed Standard*

*19;17Nov2006;14:10:24;fwfoomain01.foo.com;control;ctl;;lo0;i nbound;;;;;;;;;;;;installed EDT*

*20;17Nov2006;14:10:43;fwfoomain01.foo.com;log;drop;;hme0;
        inbound;tcp;gwt.lab.foo.com;corelinkmain01.foo.com;45;22619 ;44;3;;;;;*

*21;17Nov2006;14:10:58;fwfoogw02.foo.com;log;accept;;hme1;
        inbound;udp;corelinkmain01.foo.com;ns4.foo.net.nz;ntp-udp;ntp-udp;76;2;;;;;*

## 3.2   Figures

Figure 1 shows the help screen options available. If a user forgets to supply or a naïve user who does not have any guess about the option available to run SRr, then instead of throwing an error a message is displayed instructing the user to supply a command line input out of available options.  In figure 2 the status report generated by SRr against source attribute is displayed.  Figure 3 shows the status report generated by SRr against destination attribute. Figure 4 shows the generated status report against service attribute. Figure 5 shows the generated status report against the user defined composite attribute consisting of primitive attribute interface_name, interface_directory. Figures 6 depict the progress of status report generation and figure 7 shows the status report generated as a whole.

Figure 1. Help Screen Display

Figure 2. Status report generated for source attribute

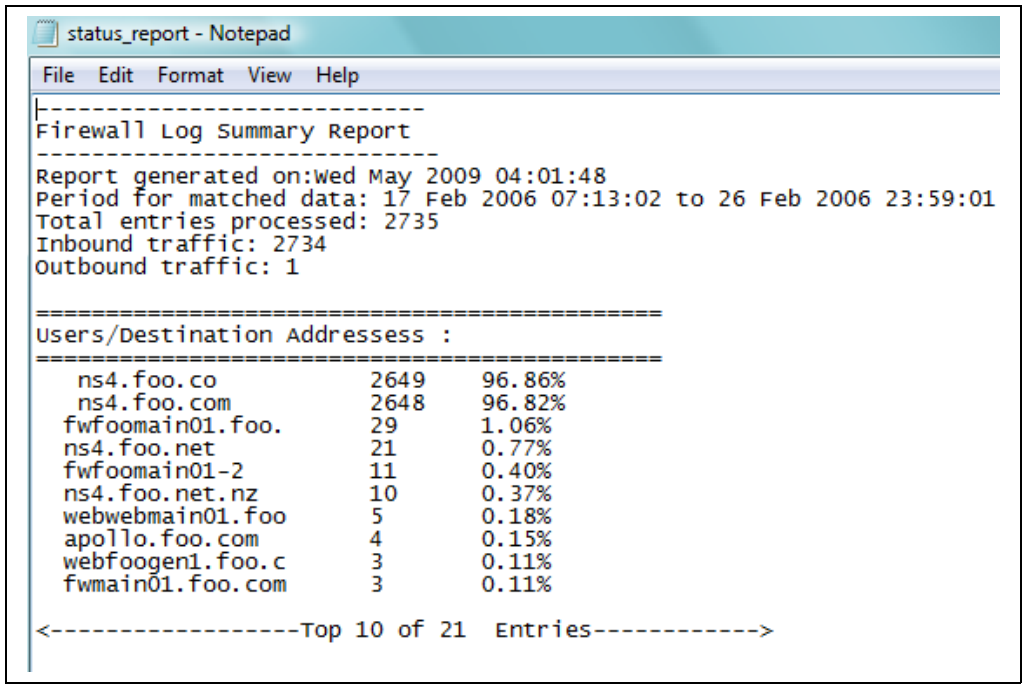

Figure 3. Status report generated for destination attribute

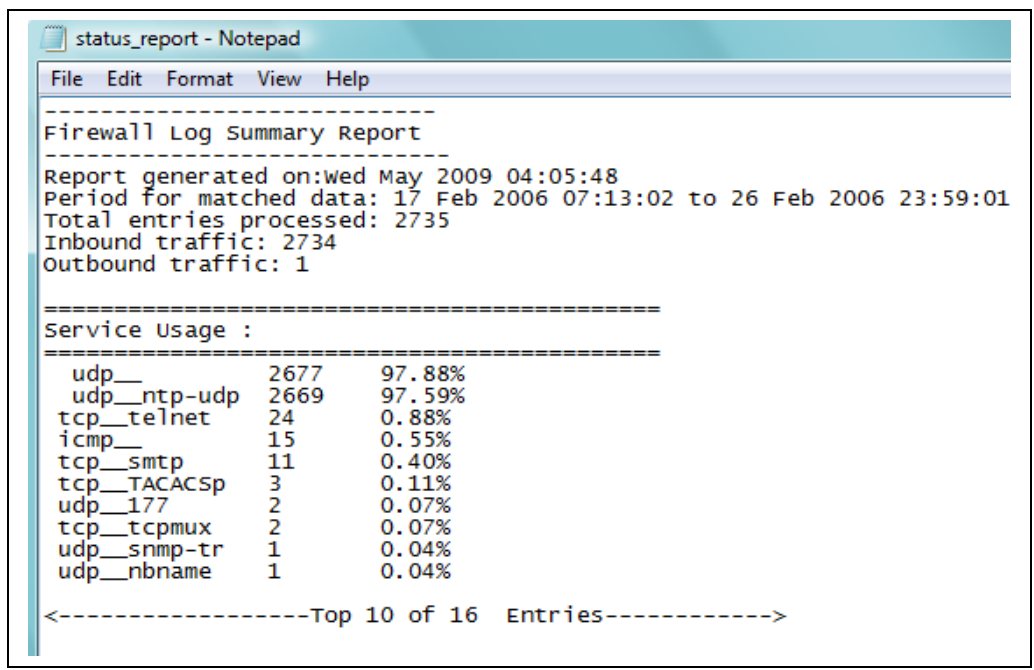

Figure 4. Status report generated for service attribute

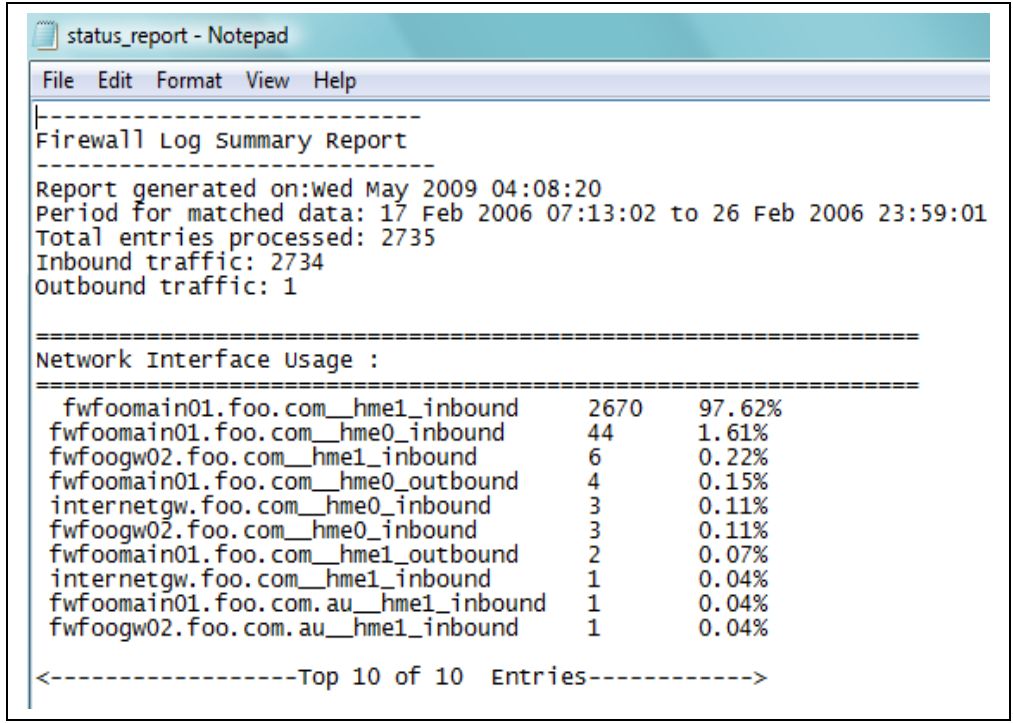

Figure 5. Status report generated for network interface attribute

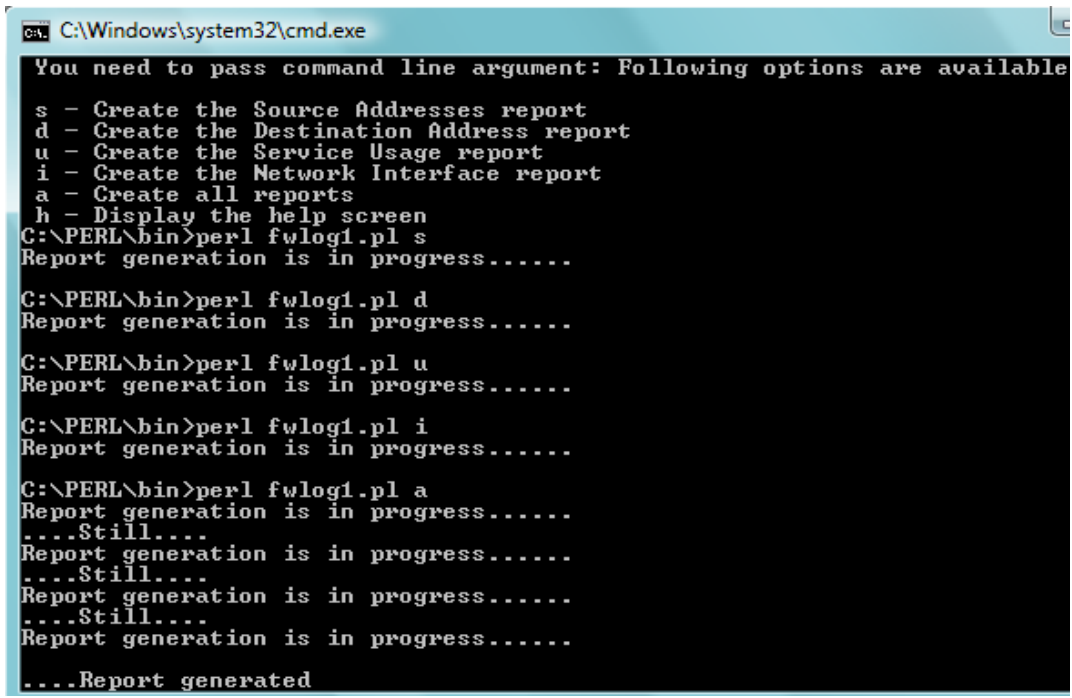

Figure 6. Screen shot showing the progress of report generation

```
Firewall Log Summary Report
============================
Report generated on:Wed May 2009 04:10:42
Period for matched data: 17 Feb 2006 07:13:02 to 26 Feb 2006 23:59:01
Total entries processed: 2735
Inbound traffic: 2734
Outbound traffic: 1

Users/Source Addressess :
=========================
   fwrtrmain01.foo.        1337      48.88%
    corelinkmain01.f       1321      48.30%
   mlink.foo.co.uk         17        0.62%
   dhcp-100-101-167        15        0.55%
   test.lab.foo.com        14        0.51%
   gwt.lab.foo.com         12        0.44%
   webwebmain01.foo        4         0.15%
   fwfoomain01.foo.        3         0.11%
   fwfoomain01-2           3         0.11%
   devel.lab.foo.co        3         0.11%

<------------------Top 10 of 18  Entries------------>

Users/Destination Addressess :
==============================
   ns4.foo.co              2649      96.86%
    ns4.foo.com            2648      96.82%
   fwfoomain01.foo.        29        1.06%
   ns4.foo.net             21        0.77%
   fwfoomain01-2           11        0.40%
   ns4.foo.net.nz          10        0.37%
   webwebmain01.foo        5         0.18%
   apollo.foo.co           4         0.15%
   webfoogeni.foo.c        3         0.11%
   fwmain01.foo.com        3         0.11%

<------------------Top 10 of 21  Entries------------>

Service Usage :
===============
   udp__                   2677      97.88%
   udp__ntp-udp            2669      97.59%
   tcp__telnet             24        0.88%
   icmp__                  15        0.55%
   tcp__smtp               11        0.40%
   tcp__TACACSp            3         0.11%
   udp__177                2         0.07%
   tcp__tcpmux             2         0.07%
   udp__snmp-tr            1         0.04%
   udp__nbname             1         0.04%

<------------------Top 10 of 16  Entries------------>

Network Interface Usage :
=========================
   fwfoomain01.foo.com__hme1_inbound      2670    97.62%
   fwfoomain01.foo.com__hme0_inbound      44      1.61%
   fwfoogw02.foo.com__hme1_inbound        6       0.22%
   fwfoomain01.foo.com__hme0_outbound     4       0.15%
   internetgw.foo.com__hme0_inbound       3       0.11%
   fwfoogw02.foo.com__hme0_inbound        3       0.11%
   fwfoomain01.foo.com__hme1_outbound     2       0.07%
   internetgw.foo.com__hme1_inbound       1       0.04%
   fwfoomain01.foo.com.au__hme1_inbound   1       0.04%
   fwfoogw02.foo.com.au__hme1_inbound     1       0.04%

<------------------Top 10 of 10  Entries------------>
```

Figure 6. Screen shot showing all status report as a whole

## 4. CONCLUSION

There are six reports generated using SRr in this research paper. The functioning of SRr is not confined till here. SRr is a customizable tool and can be applied to other applications as well where one reads an input file. The only requirement to use SRr is the input file in textual format. The use of Perl in Computer Science application is gaining immense popularity rapidly and our work is one more milestone in the same arena. We have developed SRr primarily considering System/ Network/ Database administrator's role in view. SRr provides direct benefits to those guys who read input file in textual format and want to play with that as per their requirement. Our work is among very few research works where Perl provides helping hand in System/Network/Database administration. We believe that our work will help communities, academia and industry and prove itself as life saver.

## REFERENCE

[1]. www.cpan.org

[2]. Sugam Sharma, Tzusheng Pei, and HHP Cohly, Raphael Isokpehi, and N Meghanathan(2007), "To Access PubMed Database to Extract Articles Using Perl_Su," *International Journal of Computer Science and Network Security (IJCSNS).*